\documentclass{aa}  

\usepackage{graphicx}
\usepackage{xcolor}
\usepackage{txfonts}

\newcommand{\beq}{\begin{equation}}
\newcommand{\eeq}{\end{equation}}
\newcommand{\beqa}{\begin{eqnarray}}
\newcommand{\eeqa}{\end{eqnarray}}

\usepackage{natbib}
\bibpunct{(}{)}{;}{a}{}{,} 

\begin{document}

   \title{Kink instability of triangular jets in the solar atmosphere}


%
%

   \author{T. V. Zaqarashvili
          \inst{1,4,5}
          \and
          S. Lomineishvili \inst{2,5}
         \and
          P. Leitner \inst{1}
           \and
          A. Hanslmeier \inst{1}
           \and
          P. G\"om\"ory \inst{2}
          \and
          M. Roth \inst{3}
 }

   \institute{Institute of physics, University of Graz
             Universit\"atsplatz 5, 8010 Graz, Austria\\
              \email{teimuraz.zaqarashvili@uni-graz.at}
              \and
           Astronomical Institute, Slovak Academy of Sciences, 05960 Tatransk\'a Lomnica, Slovakia
                       \and
           Leibniz-Institut f\"ur Sonnenphysik (KIS), D-79104 Freiburg, Germany
                    \and
             Ilia State University, Cholokashvili ave. 3/5, Tbilisi, Georgia
                      \and
            Abastumani Astrophysical Observatory, Mount Kanobili, Georgia \\  
             }

   \date{Received ; accepted }

 
  \abstract
   {It is known that hydrodynamic triangular jets (i. e. the jet with maximal velocity at its axis, which linearly decreases at both sides)  are unstable to antisymmetric kink perturbations. The inclusion of magnetic field may lead to the stabilisation of the jets. Jets and complex magnetic fields are ubiquitous in the solar atmosphere, which suggests the possibility of the kink instability in certain cases. }
   {The aim of the paper is to study the kink instability of triangular jets sandwiched between magnetic tubes/slabs and its possible connection to observed properties of the jets in the solar atmosphere.}
   {A dispersion equation governing the kink perturbations is obtained through matching of analytical solutions at the jet boundaries. The equation is solved analytically and numerically for different parameters of jets and surrounding plasma. The analytical solution is accompanied by a numerical simulation of fully nonlinear MHD equations for a particular situation of solar type II spicules.  }
   {MHD triangular jets are unstable to the dynamic kink instability depending on the Alfv\'en Mach number (the ratio of flow to Alfv\'en speeds) and the ratio of internal and external densities. When the jet has the same density as the surrounding plasma, then only super Alfv\'enic flows are unstable. However, denser jets are unstable also in sub Alfv\'enic regime. Jets with an angle to the ambient magnetic field have much lower thresholds of instability than field-aligned flows. Growth times of the kink instability are estimated as 6-15 min for type I spicules and 5-60 s for type II spicules matching with their observed life times. Numerical simulation of full nonlinear equations shows that the transverse kink pulse locally destroys the jet in less than a minute in the conditions of type II spicules.  }
   {Dynamic kink instability may lead to full breakdown of MHD flows and consequently to observed disappearance of spicules in the solar atmosphere.}


   \maketitle
%

\section{Introduction}

Flows and jets are essential building features of the solar atmosphere. Many different types of jets are frequently observed in the solar chromosphere/corona like spicules/mottles  \citep{Beckers1968, DePontieu2007,Ruppe2009,Tsiropoula2012,Moore2011,Sterling2020}, macrospicules \citep{Pike1998}, X-ray jets \citep{Shibata1992, Savcheva2007, Moore2013, Sterling2015,Sterling2016,Sterling2019}, EUV jets \citep{Chae1999, Zhang2014}, chromospheric anemone jets \citep{Shibata2007,Nishizuka2011} amongst others. Coronal X-ray jets can be driven by magnetic reconnection after emergence of new bipolar magnetic flux \citep{Yokoyama1995, Moore2011} or mini-filaments \citep{Sterling2015,Sterling2016, Raouafi2016}, while rebound shock waves may lead to classical spicules and macrospicules  \citep{Hollweg1982,Murawski2010,Murawski2011}.

Hydrodynamic flows are generally unstable \citep{Chandrasekhar1961,Drazin1981}, which may lead to the energy dissipation and to the consecutive heating of solar atmospheric plasma. Different flow profiles lead to the different types of instabilities. The simplest is the basic flow of two fluids in parallel infinite streams of different velocities, which is subject to Kelvin-Helmholtz instability \citep{Helmholtz, Kelvin}. Kelvin-Helmhotz instability has been intensively observed in the solar atmosphere at boundaries of rising coronal mass ejections   \citep{Ofman2011,Foullon2011,Foullon2013,Mostl2013}, in solar prominences \citep{Berger2010,Ryutova2010} and in jets \citep{Zhelyazkov2015, Zhelyazkov2018,Li2019}. Various types of flows with smooth transverse profiles are also unstable in certain conditions \citep{Drazin1981}. Another interesting process is connected to the transverse displacement of jet axis, which becomes unstable to the dynamic kink instability due to the centripetal force acting on flows in a curved path \citep{Zaqarashvili2020}. The instability may lead to the observed linear transverse motions of spicule axes \citep{DePontieu2012, Kuridze2015} in certain conditions.

Flows with smoothed transversed profiles are more difficult to be studied. The simplest case is the flow with a linear transverse profile, which can be analytically studied in various situations  \citep{Drazin2002}. A jet having maximal velocity at the axis which linearly tends to zero at boundaries is a simple model but allows relevant conclusions about jets in the solar atmosphere. This {\it triangular jet} is unstable to the antisymmetric or kink instability for long wavelength perturbations in hydrodynamics \citep{Drazin2002}. As the solar atmosphere is generally perceived by the magnetic field, the stability of triangular jets must be studied in magnetohydrodynamic (MHD) approximation. 

A magnetic field aligned with the axis of a flow generally stabilises the sub Alfv\'enic flows  \citep{Chandrasekhar1961}, while
the transverse magnetic field seems to have no effect on the instabilities \citep{Sen1963,Ferrari1981,Cohn1983}. Therefore, the magnetic
field strength (namely Alfv\'en Mach number i.e. the ratio of flow to Alfv\'en speeds) and topology are crucial for the threshold of flow instability. Flows with angles to the magnetic field, e. g. axially moving twisted magnetic flux tubes, can be always unstable \citep{Zaqarashvili2010, Zaqarashvili2014}. The magnetic filed 
in the solar atmosphere is highly inhomogeneous and has a complex topology. Therefore, the field may suppress the flow instability in some places leading to the formation of relatively stable flows. However, in some places flows may become unstable leading to the heating and turbulence of plasma.

Here we study the stability of triangular jets in the presence of a magnetic field with application to the solar atmosphere. We derive the analytical dispersion equations
for antisymmetric (kink) modes of a triangular jets imbedded in external magnetic field. The dispersion equations have complex solutions in certain conditions, which indicate to the instability of the jets. We also performed numerical simulations, which fairly confirmed the analytical thresholds and growth rates. Finally, the results were applied to type I and type II spicules in the solar atmosphere, which showed interesting coincidence of instability properties and spicule dynamics.


\begin{figure*}
\centering
   \includegraphics[width=6.4cm]{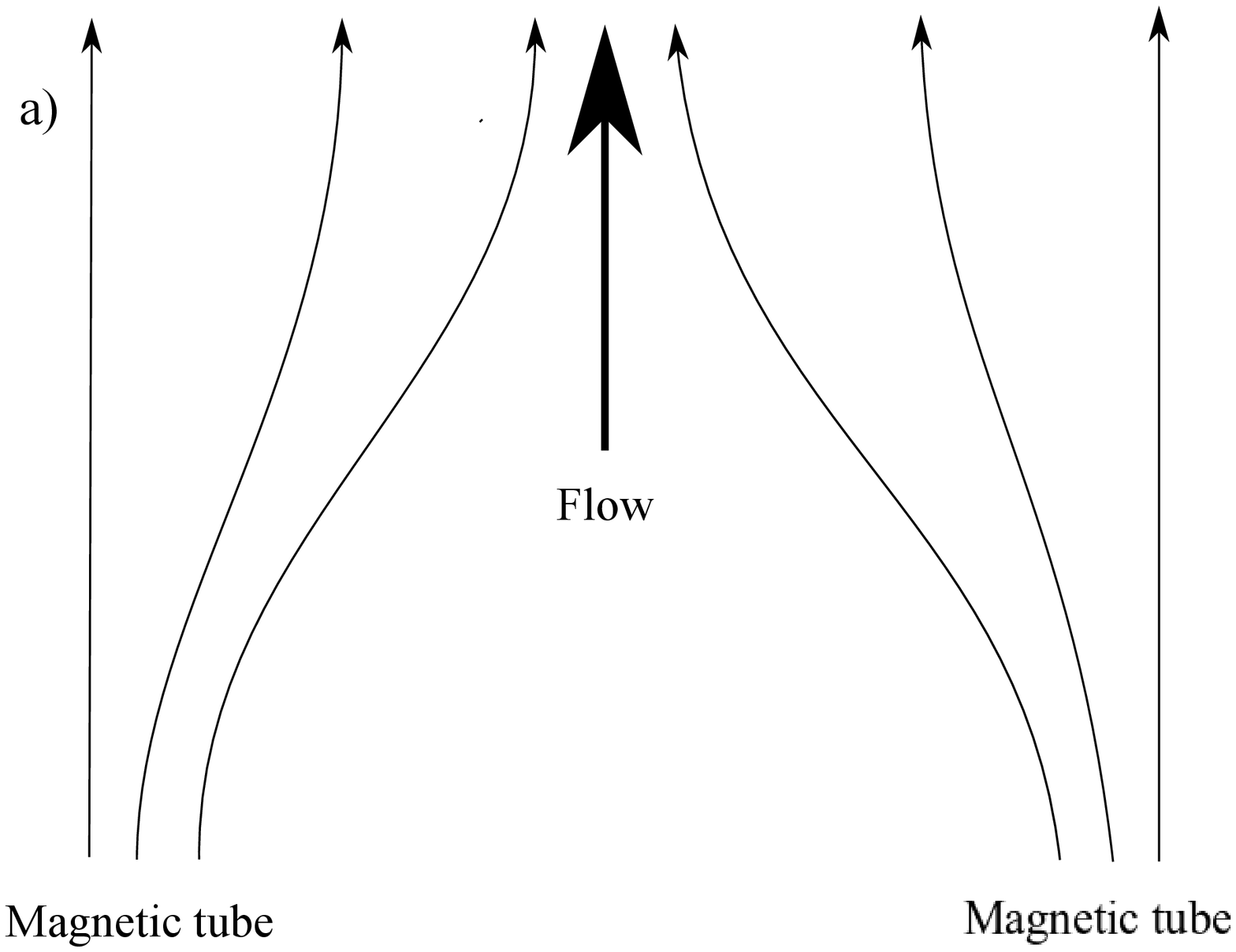}
    \includegraphics[width=5.3cm]{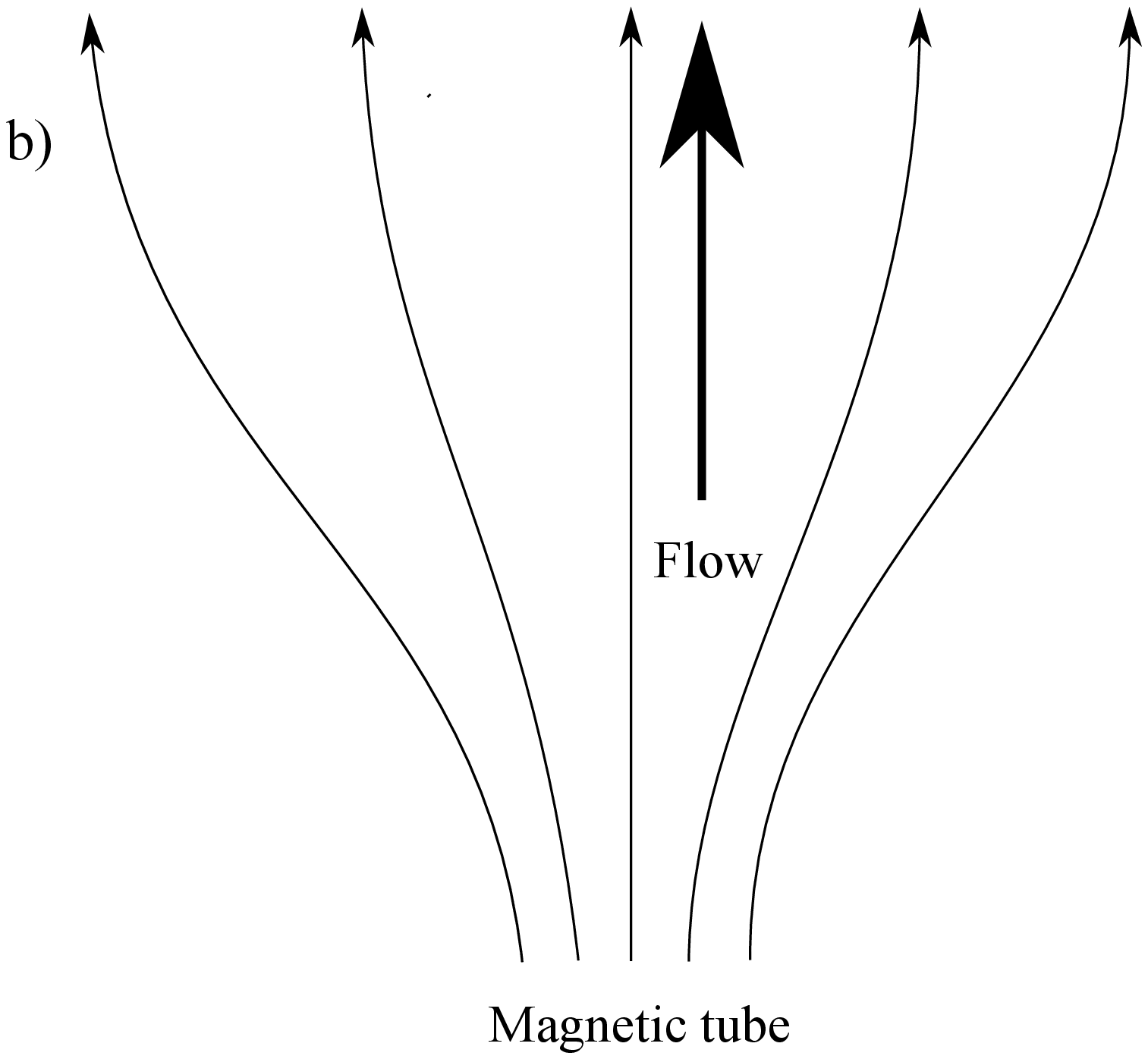}
      \includegraphics[width=6.4cm]{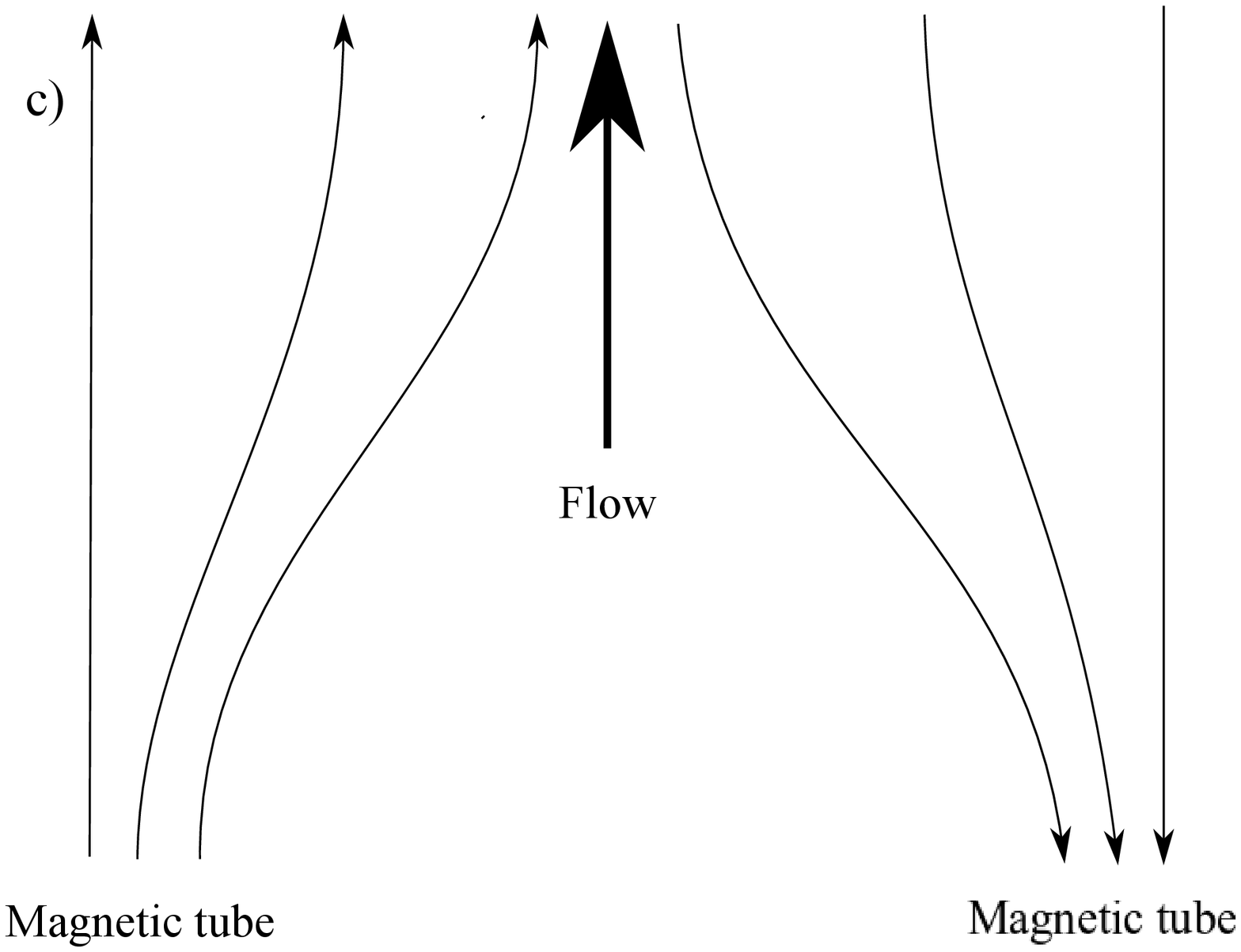}
     \caption{Possible channels of plasma flows in the solar atmosphere: a) between neighbouring magnetic flux tubes of the same polarity; b) inside a flux tube; c) between neighbouring flux tubes of opposite polarity.}
     \label{Figure1}
\end{figure*}

%
%
%

\section{Main equations}

For stability analysis of inhomogeneous jets we use simplest incompressible
approximation. Compressibility in general affects the flow stability (Sen, 1964, Gill,
1965), but the basic properties of the instability are still seen in the incompressible limit. Therefore, we consider single
fluid incompressible linearized magnetohydrodynamic (MHD) equations:
\begin{equation}
\label{eq:1} \rho \left [ {{\partial }\over {\partial t}} + {\vec
V}{\cdot}{\nabla} \right ]{\vec u} + \rho ({\vec
u}{\cdot}{\nabla}){\vec V}=-{\nabla}p_t + {1\over {4\pi}} ({\vec
B}{\cdot}{\nabla}){\vec b},
\end{equation}
\begin{equation}\label{eq:u_z}
\left [ {{\partial }\over {\partial t}} + {\vec V}{\cdot}{\nabla}
\right ]{\vec b}= ({\vec B}{\cdot}{\nabla}){\vec u}+({\vec
b}{\cdot}{\nabla}){\vec V},
\end{equation}
\begin{equation}\label{eq:3}
{\nabla}{\cdot}{\vec u}=0,\,\,\, {\nabla}{\cdot}{\vec b}=0,
\end{equation}
where $\rho$, ${\vec B}$ and ${\vec V}$ are unperturbed density,
magnetic field and flow, while ${\vec b}$, ${\vec u}$ and $p_t$ are
perturbations in the magnetic field, velocity and total
(hydrodynamic plus magnetic) pressure. The unperturbed magnetic field is assumed
to be homogeneous. Gravity effects are neglected at this stage.

We consider a Cartesian coordinate system $(x,y,z)$ and a plasma jet, which has a slab structure
along the $x$ axis with the half width of $d$ and  flows in $(y,z)$ plane. The velocity of the jet
is homogeneous with regards $y$ and $z$, but can be either homogeneous or inhomogeneous
across the slab. The unperturbed magnetic field is directed along the $z$ axis. The solutions of Eqs. (\ref{eq:1}-\ref{eq:3}) can be searched in terms of normal modes $\Psi(x)\exp i(k_y y+ k_z z-\omega t)$. Then the continuity of Lagrangian displacement and total pressure at the
slab boundaries gives the dispersion relation for the normal modes with generally complex $\omega$. Note, that the incompressible limit neglects the incoming and outgoing waves from the slab (i.e. leaky modes are absent), therefore the complex frequency means real
instability of the normal modes. 

The magnetic field of quiet Sun regions is concentrated in thin tubes at the photospheric level. However, the tubes rapidly expand upwards in the chromosphere and may merge at some heights (Fig. \ref{Figure1}). The plasma, which forms spicules at higher heights, may flow in three different channels: inside the tubes, between the neighbouring tubes with the same polarity and between the tubes with opposite polarities (shown by arrows on Fig. \ref{Figure1}). There is no firm observational evidence in which of these channels plasma flows. The different channels may support  the formation of spicules with different stability properties. The direction and speed of flows strongly depend on the formation mechanism of different jets, which is beyond the scope of present paper.  Here we assume that the jets are already formed by some mechanism and study their stability for different parameters. The stability of hydromagnetic flows depends on the transverse profile of the flow and the direction of the flow with respect to the magnetic field. Before we go to triangular jets, we briefly review the stability of homogeneous jets.

 \begin{figure}
   \centering
   \includegraphics[angle=0,width=7cm]{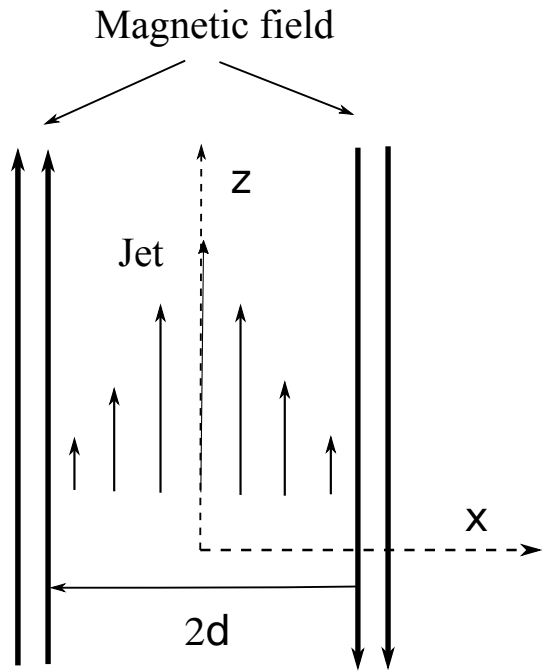}
      \caption{Sketch of considered analytical set up, which corresponds to the situation of Figure 1c. A nonmagnetic triangular jet is located between the magnetic fields of opposite polarities.
              }
         \label{Figure2}
   \end{figure}
 
\section{Homogeneous jets}

Homogeneous jets are studied elsewhere, therefore we briefly describe the main properties of its instability.
The main conclusion about the stability is that the flow-aligned magnetic field stabilises the sub Alfv\'enic flows \citep{Chandrasekhar1961,Sen1964,Ferrari1980,Hardee1988,Singh1994}. However, the
perpendicular magnetic field does not affect the instability, therefore one
always can find the exponentially growing unstable modes in a jet that has a perpendicular component to the field. 

The dispersion relation of antisymmetric perturbations for a magnetised plane jet of half-width $d$ can be written as
$$
\left [1+\frac{{\rho_0}}{{\rho_e}}\tanh(kd)  \right ]\omega^2-2\frac{{\rho_0}}{{\rho_e}}({\vec k}{\cdot}{\vec V})\tanh(kd)\omega+\frac{{\rho_0}}{{\rho_e}}\tanh(kd) ({\vec k}{\cdot}{\vec V})^2-
$$
\begin{equation}\label{slab}
\frac{{\rho_0}}{{\rho_e}}\tanh(kd)({\vec k}{\cdot}{\vec V_{A0}})^2-({\vec k}{\cdot}{\vec V_{Ae}})^2=0,
\end{equation}
where ${\rho_0}$ (${\rho_e}$) and ${\vec V_{A0}}$ (${\vec V_{Ae}}$)
are the density and Alfv\'en speed  inside (outside) the jet,
${\vec k}$ is the wave vector in $yz$-plane. This equation can be obtained from the general dispersion relation of \citet{Singh1994} as an incompressible limit. 
If the jet is directed along the unperturbed magnetic field (i.e. $\vec V \parallel \vec V_A$), where the Alfv\'en speed is assumed to have the same strength and direction inside and outside the slab for simplicity, then phase speed is always real for any sub Alfv\'enic flows ($\vec V < \vec V_A$). However, if the jet has an angle with the magnetic field, then the harmonics
perpendicular to the magnetic field i.e. with ${\vec k}{\cdot}{\vec V_{A}}=0$ always have complex phase speed. Therefore, the jet, which flows at an angle
to the magnetic field, is always unstable for antisymmetric perturbations. This antisymmetric perturbations mean the transverse displacement of the jet axis like as kink waves in magnetic slabs \citep{Edwin1982}, hence they are unstable to dynamic kink instability \citep{Zaqarashvili2020}.

The jet flowing inside the magnetic flux tube (the middle arrow in Fig. \ref{Figure1}) may probably follow the magnetic field. Then, it can be stable against dynamic   kink instability and may represent the classical spicules. However, the magnetic fields can be inclined at the tube boundaries, and thus make an angle with respect to upward flows. Therefore, the flows between the tubes may have the component perpendicular to the magnetic field and then they can be unstable due to the the dynamic kink instability.

\section{Triangular jets}
\label{sec:inhomogeneous-jet}

Now we consider a jet, which has transverse inhomogeneity in velocity, and study how this transverse structure affects the instability properties of the jet.
For simplicity, we assume simplest linear profile of the flow and consider a triangular jet, which has maximum velocity at the slab axis and linearly decreases towards slab boundaries. The stability of the jet is well studied in nonmagnetic fluids. It was shown that the jet is generally unstable for long wavelength antisymmetric normal modes, while it is generally stable for the symmetric modes \citep{Drazin2002}. It is of our interest to study how an external magnetic filed influences the stability properties of the jet. Analytical solution of MHD triangular jets is complicated, therefore we consider the situation when a nonmagnetic jet is sandwiched between two magnetic environments (Fig. \ref{Figure2}). Especially interesting situation may arise when a jet flows between the two tubes with opposite polarities (Figure 1c), where a neutral sheet is formed. The magnetic field becomes negligible between the tubes and the nonmagnetic jet is a good approximation.

Therefore, for the triangular jet we consider
\beqa |B_z|=const,\,\, V=0, \,\, &{\rm for}& \,\, |x| > d ,\,\,
\label{eq:I}
\\
B_z=0,\,\, {\vec V}={\vec V_0}\left (1-\alpha {{|x|}\over d} \right ), \,\,
&{\rm for}& \,\, |x|<d, \,\, \label{eq:II} \eeqa
where the flow, $\vec V$, has $y$ and $z$ components. The flow velocity is maximal at the slab axis and linearly decreases towards boundaries. The  parameter $0 \leq \alpha \leq 1$ governs the rate of flow inhomogeneity. $\alpha=0$ means the homogeneous jet, which recovers the situation of previous subsection, while $\alpha=1$ describes the jet which tends to zero at the slab boundaries. $\alpha=1$ case was considered by \citet{Drazin2002}, but with magnetic field free environment. The flow density, $\rho_0$, in general, is different
from the surrounding plasma i.e $\rho_0 \not= \rho_e$ ($\rho_e$ is the density in external medium).

Then Eqs. (\ref{eq:1}-\ref{eq:3}) lead to the equations
\begin{equation}\label{eq:eq1}
\rho({\vec k}{\cdot}{\vec V}-\omega)\left [{{\partial^2 }\over
{\partial x^2}} - k^2 \right ]u_x={k_zB_z\over {4\pi}}\left
[{{\partial^2 }\over {\partial x^2}} - k^2 \right ]b_x,
\end{equation}
\begin{equation}\label{eq:eq2}
({\vec k}{\cdot}{\vec V}-\omega)b_x=k_zB_zu_x,
\end{equation}
where $k=\sqrt{k^2_y+k^2_z}$.

When $\omega \not= {\vec k}{\cdot}{\vec V}$ and $\omega \not= k_z V_A$, then the plasma dynamics inside and outside of the slab is governed by
the equation
\begin{equation}\label{eq:gen}
\left [{{\partial^2 }\over {\partial x^2}} - k^2 \right ]u_x=0.
\end{equation}
The solution of this equation is a combination of exponential functions and it depends on boundary conditions on the slab center, boundaries and infinity. 

We require that the solution vanishes at infinity outside the slab, therefore the resulted expression is 
\beqa u_x=Ae^{-k(|x|-d)},\,\, &{\rm for}& \,\,
|x|>d. \label{eq:sol+} 
\eeqa
The solutions inside the slab can be antisymmetric (sinuous) or symmetric (varicose) with regards to the slab center. 

For the antisymmetric kink mode the solution inside the jet is \citep{Drazin2002}
\beqa 
u_x=B{{\cosh{k x}}\over {\cosh{k
d}}}+D{{\sinh{k |x|}}\over {\sinh{k d}}},\,\, &{\rm for}& \,\,
|x|<d.\,\, \label{eq:solutII} 
\eeqa
%

These solutions must satisfy the continuity of Lagrangian transverse velocity and total pressure at the slab boundaries, which give the equations:
 \beqa {{u_x}\over
{{\vec k}{\cdot}{\vec V}-\omega}}=const,\,\, &{\rm at}& \,\, x =
{\pm} d, \label{eq:boundary1}
 \eeqa
$$
\rho\left [{\vec k}{\cdot}{\vec V}-\omega-{{k_z^2v^2_A}\over {{\vec
k}{\cdot}{\vec V}-\omega}} \right ]{{\partial {u_x}}\over {\partial
x}}- \rho {\vec k}{\cdot}{\vec V}^{\prime}\left [1-
{{k_z^2v^2_A}\over {({\vec k}{\cdot}{\vec V}-\omega)^2}}\right
]u_x=
$$
\beqa  =const,\,\, &{\rm at}& \,\, x = {\pm} d. \label{eq:boundary2}
\eeqa

Additionally, the third equation is obtained from the pressure continuity condition at the slab center:
\beqa 
\rho\left [{\vec k}{\cdot}{\vec V}-\omega \right ]{{\partial {u_x}}\over {\partial
x}}- \rho {\vec k}{\cdot}{\vec V}^{\prime}u_x=const,\,\, &{\rm at}& \,\, x = 0. \label{eq:boundary3}
\eeqa




 \begin{figure}
   \includegraphics[angle=0,width=10cm]{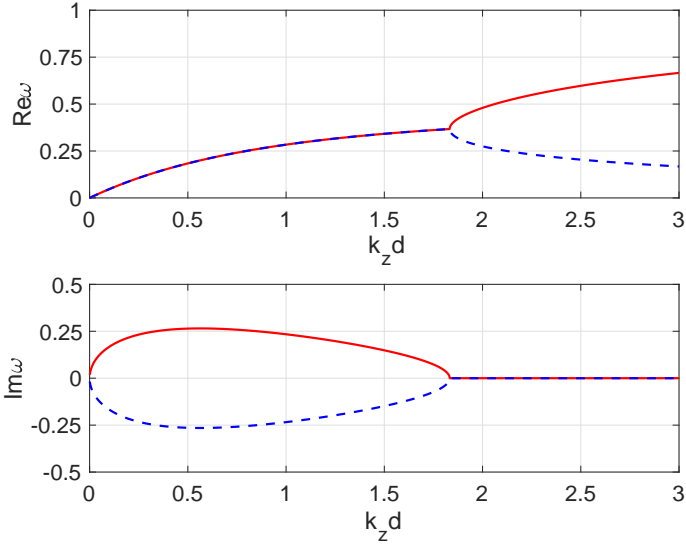}
      \caption{Solutions of antisymmetric (kink) mode (Eq. \ref{eq:disp}) for nonmagnetic case ($B_z=0$) when $\alpha=1$ and $k_y, V_{0y}=0$ (it corresponds to the solution of \citet{Drazin2002}). Upper  and lower panels show real and imaginary parts of frequency, respectively. Red solid lines correspond to unstable modes and blue dashed lines to the damped modes. Imaginary solutions (and hence instability) vanish for $k_z d > 1.8$ and for $k_z d \rightarrow 0$. Frequency and growth rate are normalised by $k_z V_{0z}$.
              }
         \label{Figure3}
   \end{figure}
 Eqs.  (\ref{eq:sol+}-\ref{eq:solutII})  and  (\ref{eq:boundary1}-\ref{eq:boundary3}) give
the following dispersion relation for the antisymmetric mode
$$
\left [1+{{\rho_0}\over {\rho_e}}\tanh{k d}\right ]\omega^3+{\vec
k}{\cdot}{\vec V_0}\left [\alpha{{\tanh{k d}}\over {k d}}-{{\rho_0}\over
{\rho_e}}(3-2\alpha)\tanh{k d}-1 \right ]\omega^2
$$
$$
+({\vec k}{\cdot}{\vec V_0})^2\Bigg [{{\alpha^2}\over {k d}}{{\rho_0 }\over
{\rho_e}}\left ( 1 - {{\tanh{k d}}\over {kd}}\right )+ {{\rho_0 }\over
{\rho_e}}(1-\alpha)(3-\alpha)\tanh{k d}
$$
$$
-\frac{k_z^2V^2_A}{({\vec k}{\cdot}{\vec V_0})^2}
\Bigg ]\omega +{{\rho_0 }\over
{\rho_e}}(1-\alpha)({\vec k}{\cdot}{\vec V_0})^3\Bigg [\frac{\alpha^2}{k^2d^2}\tanh{k d} -\frac{\alpha^2}{kd} 
$$
\begin{equation}\label{eq:disp}
-(1-\alpha)\tanh{k d} \Bigg ]+{\vec k}{\cdot}{\vec V_0} k_z^2V^2_A \left [1-
\alpha{{\tanh{k d}}\over {kd}}\right  ]=0.
\end{equation}

This is a general dispersion relation, which can be analysed in different context.

\subsection{Non-magnetised jets}

If the magnetic field is negligible ($V_A \approx 0$), then one can consider the jet in the $XZ$ plane (i. e. $k_y, V_{0y}=0$) and Eq. (\ref{eq:disp}) is transformed (for $\alpha=1$) into the dispersion equation of non-magnetised triangular jet  \citep[see][Equation~8.43]{Drazin2002}
$$
\left [1+{{\rho_0}\over {\rho_e}}\tanh{k_z d}\right
]\omega^2+k_zV_{0z}\left [{{\tanh{k_z d}}\over {k_z
d}}-{{\rho_0}\over {\rho_e}}\tanh{k_z d}-1 \right ]\omega+
$$
\begin{equation}\label{eq:disp1}
{{k_z^2 V_{0z}^2}\over {k_z d}}{{\rho_0 }\over {\rho_e}}\left ( 1 -
{{\tanh{k_z d}}\over {k_z d}}\right ) =0.
\end{equation}
This equation has complex solutions in the interval $0<k_y d < 1.8$ i.e. the flow is unstable for long wavelength perturbations.  The solution of the Eq.  (\ref{eq:disp}) in this case is shown on   Fig. \ref{Figure3}. Frequency has an imaginary part in the interval $0<k_z d < 1.8$, which vanishes for $k_z d > 1.8$.

If the magnetic field is presented, but the flow is not parallel to the magnetic field (i.e. $V_{0y} \not = 0$), then the modes with $k_y \not =0$ and $k_z=0$ (which are perpendicular to the magnetic field) are governed by the same equation Eq.  (\ref{eq:disp1}), but $k_z$ and $V_{0z}$ are replaced by $k_y$ and $V_{0y}$. Hence, the flow is always unstable for long wavelength perturbations even in the presence of magnetic field. This confirms the previous results that the normal-to-flow magnetic field does not affect the Kelvin-Helmholtz  instability \citep{Chandrasekhar1961,Sen1964,Zaqarashvili2010,Zaqarashvili2014}.

%

\subsection{$\alpha=0$ (homogeneous jet) limit}

Next we consider the limit of $\alpha=0$, which actually means a homogeneous jet. Then the Eq. (\ref{eq:disp}) is transformed into the equation 
$$
\Bigg (\left [1+\frac{{\rho_0}}{{\rho_e}}\tanh(kd)  \right ]\omega^2-2\frac{{\rho_0}}{{\rho_e}}({\vec k}{\cdot}{\vec V_0})\tanh(kd)\omega+\frac{{\rho_0}}{{\rho_e}}\tanh(kd) ({\vec k}{\cdot}{\vec V_0})^2-
$$
\begin{equation}\label{eq:disp2}
k^2_z V^2_A \Bigg )(\omega-{\vec k}{\cdot}{\vec V_0})=0,
\end{equation}
When $\omega \not= {\vec k}{\cdot}{\vec V_0}$, then we recover Eq.  (\ref{slab}) (with $\vec V_{A0}=0$). We note that Eq.  (\ref{slab}) is transformed into the dispersion relation of Alfv\'en surface waves in static magnetic slab (${\vec V_0}=0$) \citep[see][Equation~12]{Edwin1982}. The antisymmetric modes are  unstable when
\begin{equation}\label{eq:disp3}
\frac{({\vec k}{\cdot}{\vec V_0})^2 }{k^2_z V^2_A} >\frac{\rho_e+\rho_0\tanh(kd)  }{ \rho_0\tanh(kd)}.
\end{equation}
When $k_y=0$ and $\rho_e=\rho_0$, then this inequality is replaced by $V_{0z}/V_A>\sqrt{2/(1-e^{-2kd})}$, which actually means that the flow is stabilised for all wavelength perturbations when $V_A/V_{0z}>\sqrt{0.5} \approx 0.707$. When $V_A/V_{0z}< \sqrt{0.5}$, then long wavelength perturbations are stable, while the short wavelength perturbations are unstable. The critical wavenumber can be estimated as $k_z d = \ln{\left (1-2V^2_A/V^2_{0z} \right )^{-1/2}}$.

\begin{figure}
   \includegraphics[angle=0,width=10cm]{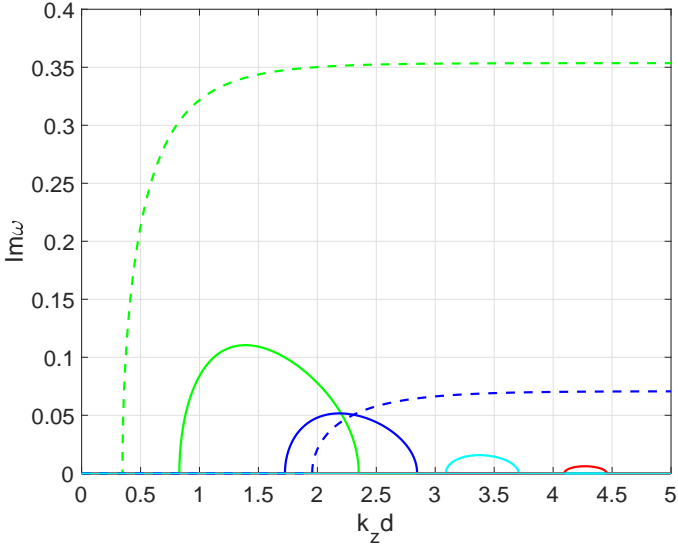}
      \caption{Growth rate (imaginary part of frequency) of antisymmetric (kink) mode vs normalised wavenumber $k_z d$ calculated from Eq.  (\ref{eq:disp}) when  $\alpha=1$ and $k_y, V_{0y}=0$ for different ratios of   inverse Alfv\'en Mach number, $M^{-1}_A$. Green, blue, cyan and red solid lines correspond to $M^{-1}_A=$0.5, 0.7, 0.9 and 1, respectively. Dashed lines show the corresponding solutions for homogeneous jet $\alpha=0$, which is already zero for $M^{-1}_A=$ 0.9 and 1. Here $\rho_0/\rho_e=1$ and the growth rate is normalised by $k_zV_{0z}$.
              }
         \label{Figure4}
   \end{figure}

\subsection{$\alpha=1$ limit}

Next we consider the limit when the jet velocity vanishes at the slab boundaries ($x=\pm d$) i.e. $\alpha=1$. In this case, Eq.  (\ref{eq:disp}) leads to the equation (see also \citet{zaqarashvili2011})
$$
\left [1+{{\rho_0}\over {\rho_e}}\tanh{k d}\right ]\omega^3+{\vec
k}{\cdot}{\vec V_0}\left [{{\tanh{k d}}\over {k d}}-{{\rho_0}\over
{\rho_e}}\tanh{k d}-1 \right ]\omega^2+
$$
$$
({\vec k}{\cdot}{\vec V_0})^2\left [{1\over {k d}}{{\rho_0 }\over
{\rho_e}}\left ( 1 - {{\tanh{k d}}\over {kd}}\right ) -\frac{k_z^2V^2_A}{({\vec k}{\cdot}{\vec V_0})^2}
\right ]\omega +
$$
\begin{equation}\label{eq:disp4}
{\vec k}{\cdot}{\vec V_0} k_z^2V^2_A \left [1-
{{\tanh{k d}}\over {kd}}\right  ]=0.
\end{equation}

\begin{figure}
   \includegraphics[angle=0,width=10cm]{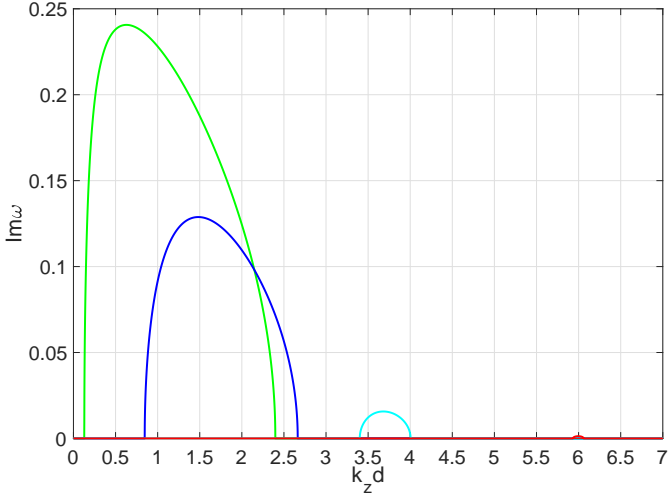}
      \caption{Same as on Fig. \ref{Figure4}, but for the density ratio of $\rho_0/\rho_e$=10. Green, blue, cyan, and red solid lines correspond to $M^{-1}_A=$0.5, 1, 2, and 2.5 respectively. Growth rate is normalised by $k_z V_{0z}$.  
              }
         \label{Figure5}
   \end{figure}
This equation is transformed into Eq.  (\ref{eq:disp1}) for non-magnetised case. 

Fig. \ref{Figure4} shows the growth rates of unstable modes against normalised wavenumber calculated from the dispersion relation (\ref{eq:disp}) for $\alpha=1$ (solid lines) for different values of Alfv\'en Mach number, $M_A=V_{0z}/V_A$ (the ratio of the flow speed at the slab center and the external Alfv\'en speed), for $\rho_0/\rho_e$=1. It is seen that the flow is unstable only for finite interval of the wave numbers. Therefore, only the harmonics with certain wave lengths are unstable. 
On the other hand, homogeneous flows ($\alpha=0$, dashed lines) are unstable when the wavelength is larger than the critical one. 
Sufficiently strong magnetic field  suppresses the instability when $M^{-1}_A>1$ for triangular jets, which means that sub-Alfv\'enic flows $V_{0z}<V_A$ are stable. But homogeneous jets can be stabilised even for slightly super Alfv\'enic regime, though they have relatively stronger growth rates.

\begin{figure}
   \includegraphics[angle=0,width=10cm]{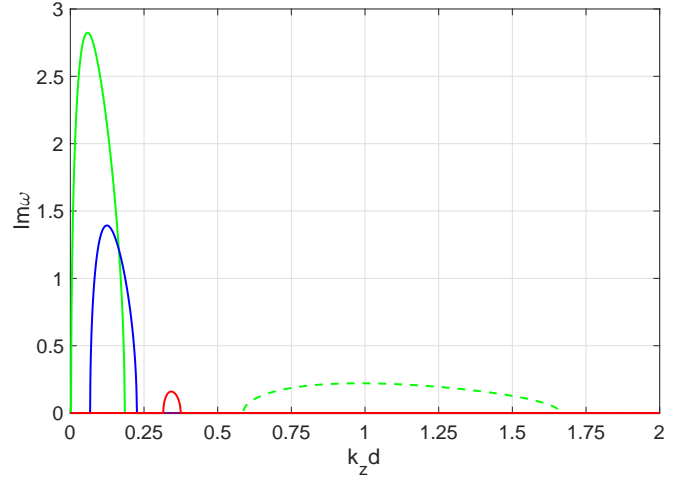}
      \caption{Growth rate of antisymmetric (kink) mode vs wavenumber for modes with two different propagation angles, $k_y/k_z$, when flow is directed with 45 degree to the magnetic field, $V_{0y}/V_{0z}=1$. Green dashed line corresponds  to $M^{-1}_A=1$ for $k_y/k_z$=1, while green, blue, and red solid lines correspond to $M^{-1}_A=$1, 5, and 10 respectively, when $k_y/k_z$=10. 
              }
         \label{Figure6}
   \end{figure}

Fig. \ref{Figure5} displays the growth rates for different values of Alfv\'en Mach number for the denser jet,  $\rho_0/\rho_e$=10. It shows that the denser jet needs stronger external Alfv\'en speed  to stabilise the kink instability (in this case Alfv\'en speed needs to be 2.5 higher than the flow speed at the slab center). It is clear from physical point of view that the denser jet has stronger kinetic energy and therefore the stronger magnetic energy it required for stabilisation.

If the flow is directed with some angle to the magnetic field, i.e. $V_{0y} \neq 0$, then the instability is not completely suppressed by the magnetic field as it was mentioned in the subsection 4.1. Fig. \ref{Figure6} shows the growth rates of unstable modes against wavenumber for $V_{0y}=V_{0z}$, i.e. the flow is directed with 45 degree to the magnetic field. In this case, the modes with the propagation angle of 45 degree to the magnetic field, $k_y/k_z=1$ (dashed line), are only stabilised by sufficiently strong magnetic field. But the modes with the propagation angle close to 90 degree to the magnetic field ($k_y/k_z$=10, solid lines) are unstable even for very high Alfv\'en speed of $V_A/V_{0z}=10$.


\begin{figure*}
   \includegraphics[angle=0,width=9cm]{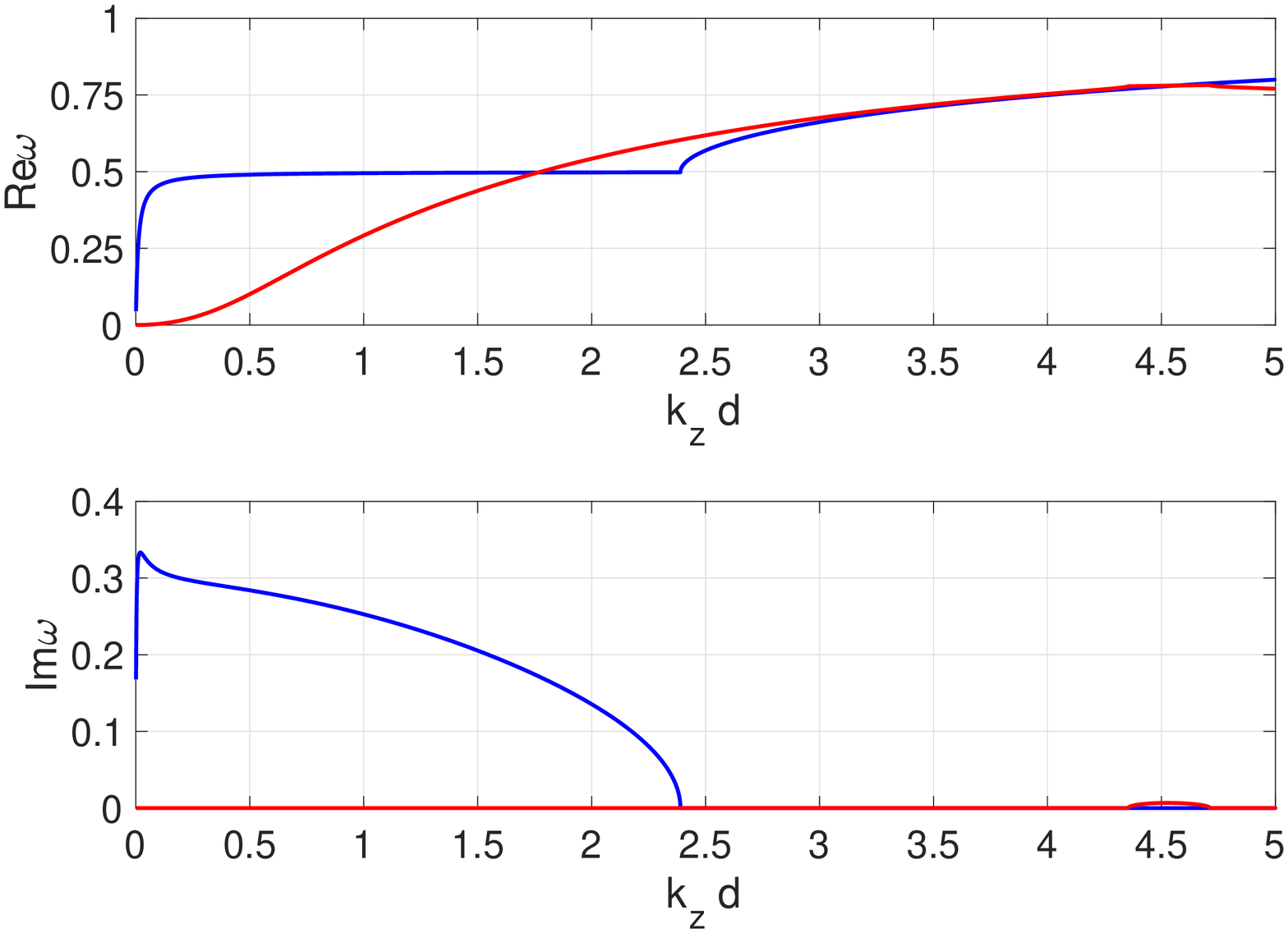}
   \includegraphics[angle=0,width=9cm]{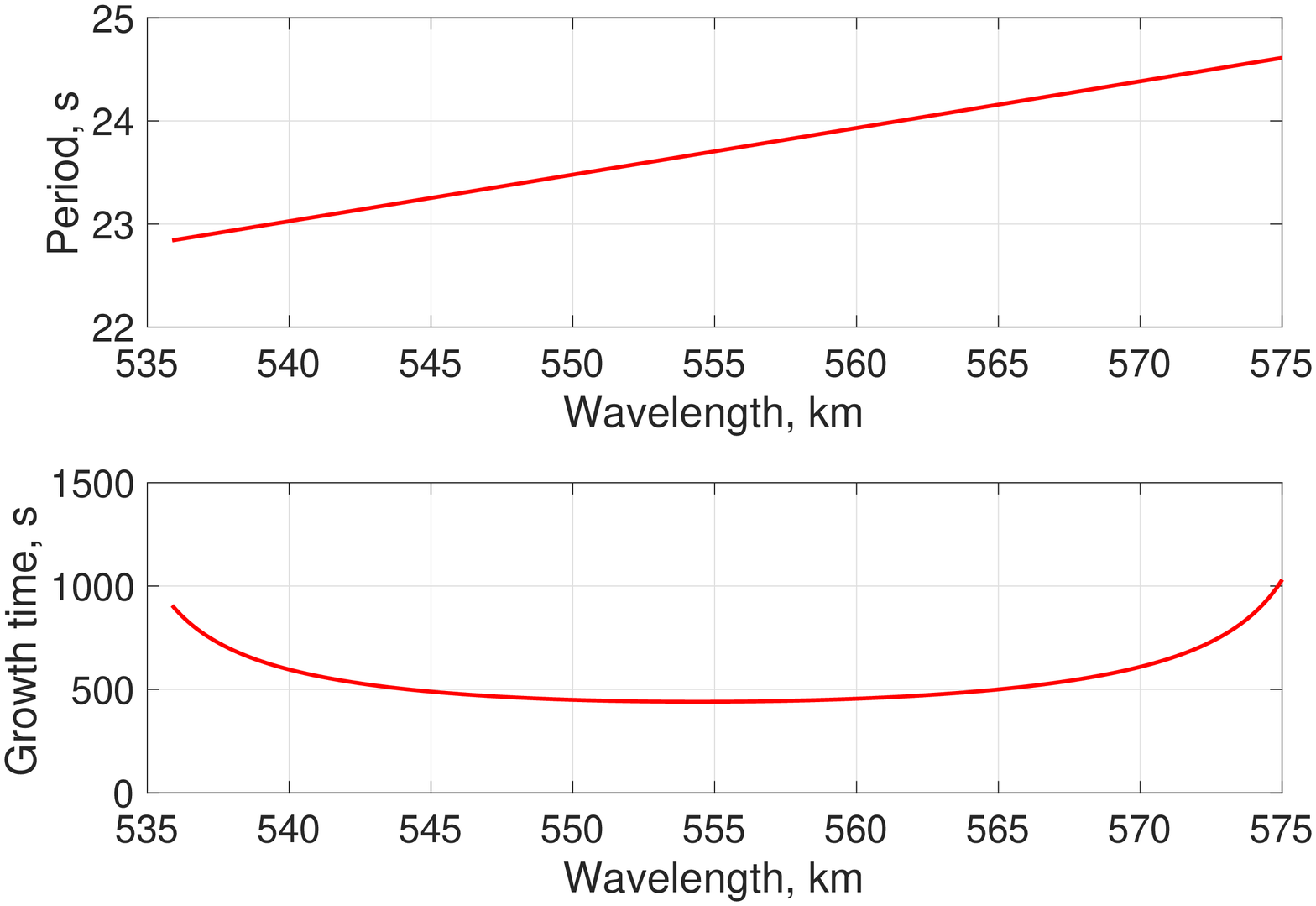}
      \caption{Solutions of antisymmetric (kink) mode (Eq. \ref{eq:disp}) in the conditions of type I spicules: $\rho_0/\rho_e=100$, $d=400$ km, $V_0=30$ km/s.  Left panels show the real (upper) and the imaginary (lower) parts of frequencies, which are normalised by $k_z V_{0z}$. Blue lines show the purely hydrodynamic case i.e.  $V_A=0$. Red lines show the case when the external Alfv\'en speed is $V_A$=200 km/s. Right panels show the periods (upper) and growth times (lower) of unstable harmonics vs wavelength in the interval of $k_z d=$4.4-4.7. Here $\alpha=1$ and $k_y, V_{0y}=0$.
              }
         \label{Figure7}
   \end{figure*}

%
%
%
%
\begin{figure*}
   \includegraphics[angle=0,width=9cm]{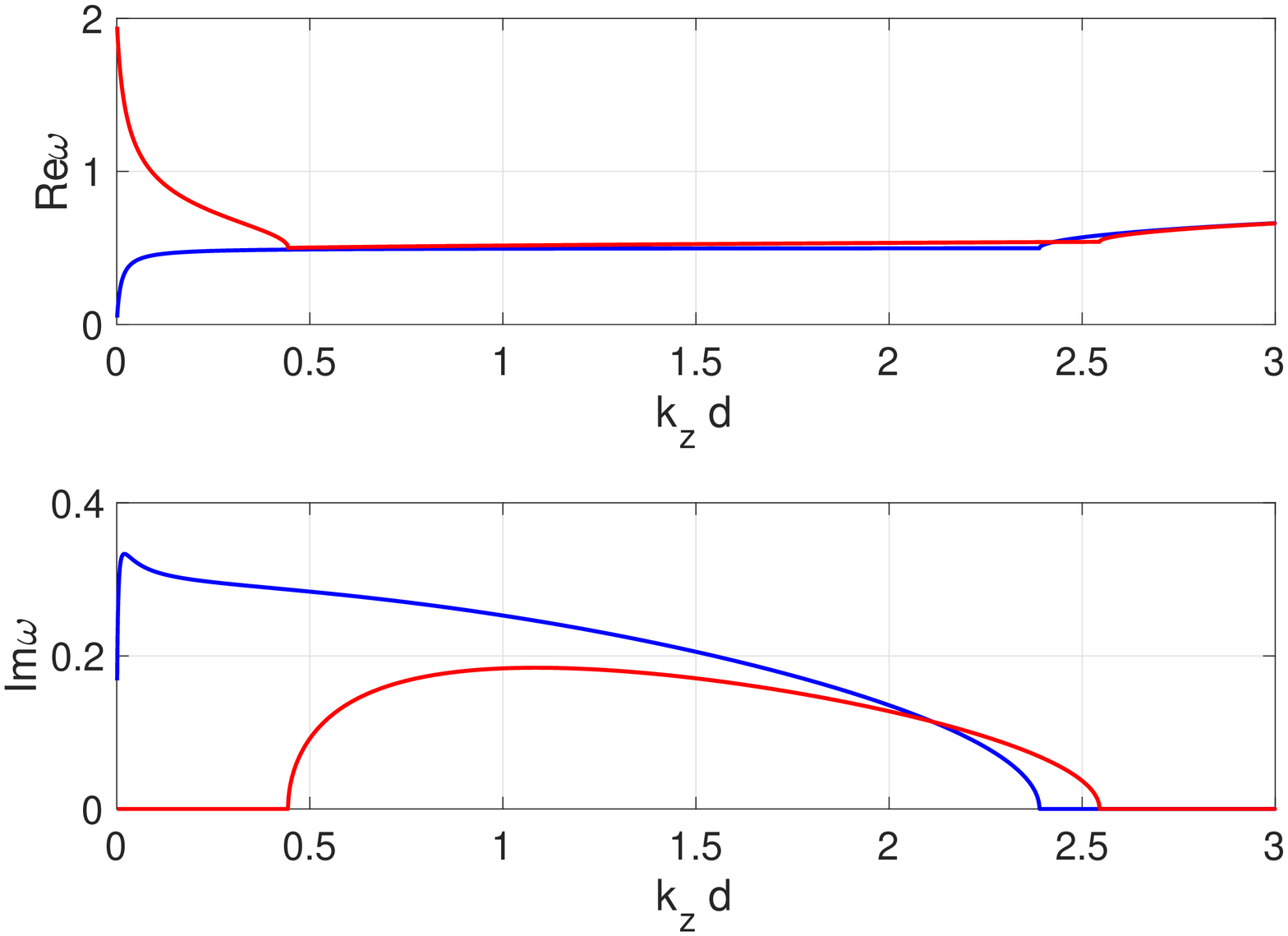}
    \includegraphics[angle=0,width=9cm]{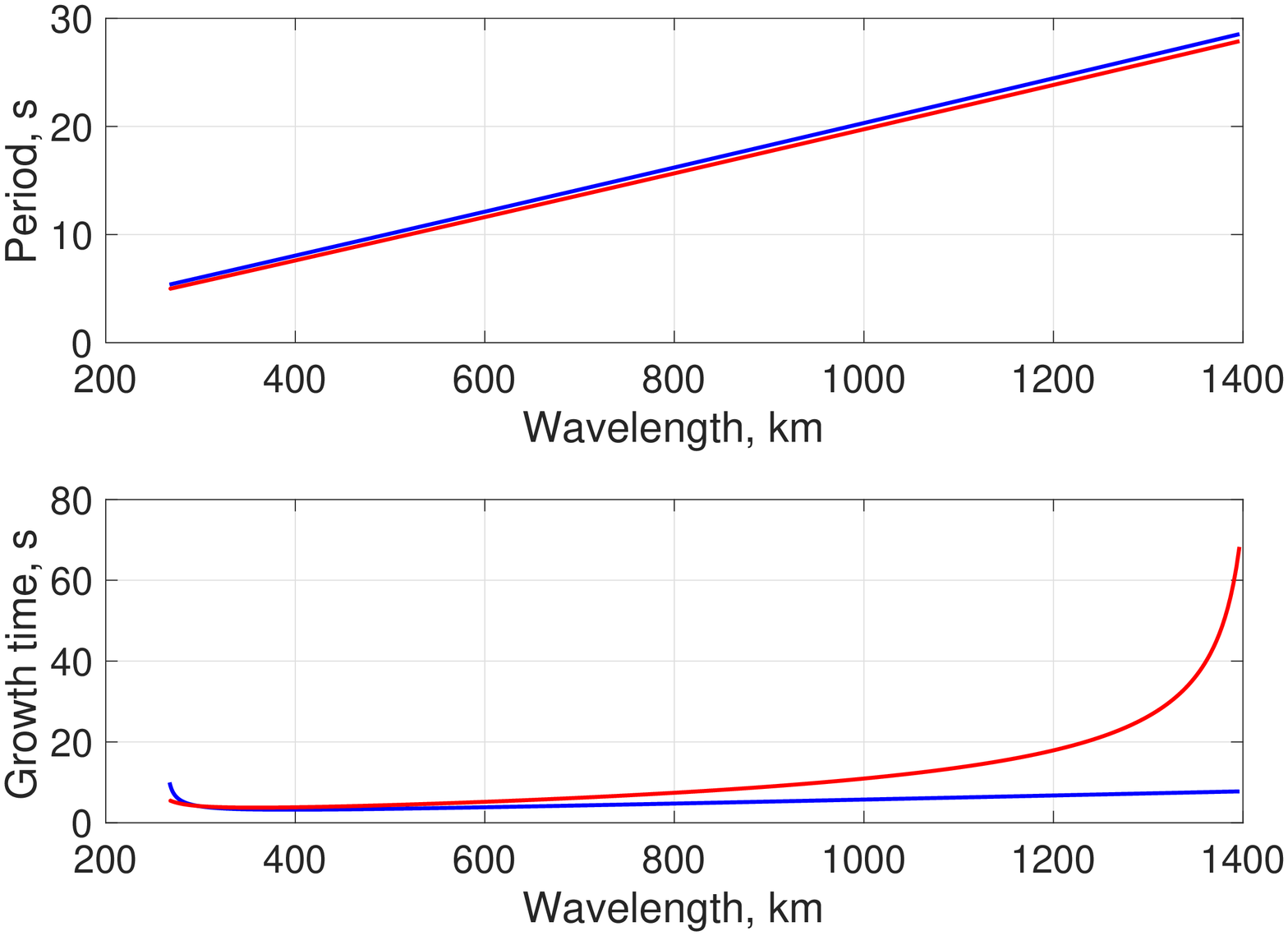}
      \caption{Solutions of antisymmetric (kink) mode (Eq. \ref{eq:disp}) in the conditions of type II spicules: $\rho_0/\rho_e=100$, $d=100$ km, $V_0=100$ km/s.  Left panels show the real (upper) and the imaginary (lower) parts of frequencies, which are normalised by $k_z V_{0z}$. Blue lines show the purely hydrodynamic case i.e.  $V_A=0$. Red lines show the case when the external Alfv\'en speed is $V_A$=200 km/s. Right panels show the periods and growth times of unstable harmonics vs wavelength in the interval of $k_z d=$0.5-2.5. Here $\alpha=1$ and $k_y, V_{0y}=0$.
              }
         \label{Figure8}
   \end{figure*}

\section{Discussion}

It was shown by \citet{Drazin2002} that hydrodynamic 2D triangular jets are unstable to antisymmetric kink modes, while they are stable to symmetric sausage modes. The kink modes are unstable in certain interval of wavelengths, while short and long wavelength perturbations are stable (see Fig. \ref{Figure3}). Here we studied the influence of magnetic field on the kink instability of triangular jets in the connection to solar atmospheric physics. In order to obtain analytical dispersion equations, we considered a triangular jet sandwiched between magnetic field tubes (or slabs), which allowed us to solve the linearised problem and led to analytical dispersion equation  (Eq. \ref{eq:disp}) using appropriate boundary conditions. 

The dispersion equation has solutions in the form of complex wave frequency, which means instability of corresponding wave modes. The kink instability occurs in the certain interval of wave numbers depending on flow speed at the axis, or more precisely on the ratio of flow to Alfv\'en speeds (Alfv\'en Mach number). Decreasing of Alfv\'en Mach number leads to the shortening of the instability interval and the shifting of the interval to higher wave numbers (see Fig. \ref{Figure4}). After certain Alfv\'en Mach number the instability is terminated, hence the magnetic field stabilises the kink instability. For the density ratio of $\rho_0/\rho_e=1$, the instability is ceased when $M^{-1}_A>1$ i. e. for sub Alfv\'enic flows (see Fig. \ref{Figure4}). While for the denser jet, $\rho_0/\rho_e=10$, the instability is ceased when $M^{-1}_A>2.5$ (see Fig. \ref{Figure5}). Physically it can be understood as follows. Kink instability is related with the centripetal force, which is generated when the axis of jet is transversally displaced so the plasma flows in a curved trajectory. This force is directed in direction of transverse displacement and hence tries to enhances the displacement \citep{Zaqarashvili2020}. On the other hand, the Lorentz force tries to straighten the magnetic field lines and hence acting against the instability. Denser jet has more inertia and therefore stronger magnetic field (and consequently lower Alfv\'en Mach number) is needed to stabilise the kink instability. 

If the triangular jet is directed with an angle to the ambient magnetic field, then the instability may start well below the threshold (see Fig. \ref{Figure6}). This situation may arise for inclined jets when the neighbouring magnetic field lines are vertical (as on Figure 1a, 1c) or for vertical jets when the surrounding tubes are twisted. In either cases, the sub Alfv\'enic jets could be unstable to the dynamic kink instability. This process is related with the known result that only flow-aligned component of magnetic field stabilises the flow instability \citep{Chandrasekhar1961}. The magnetic field component across the flow allows to the velocity field to grow parallel to magnetic field lines without influence from the Lorentz force, hence the flow instability starts in sub Alfv\'enic regimes.

The antisymmetric kink instability discussed here can be used to model the instability of various types of jets in the solar atmosphere. Here we consider the stability of spicules as it is seen under obtained results. It must be mentioned, however, that the real conditions of the solar atmosphere (magnetic field structure, jet density structure, etc.) are much more complicated comparing with simplified set up of this paper. Therefore, the results show only general properties of jet instability, which can significantly vary for different types of jets and solar atmospheric conditions. 
 
\subsection{Comparison with solar observations: instability of spicules}

Spicules are chromospheric plasma jets flowing upwards into the lower corona, therefore they are almost 100 times denser and cooler than the corona. Spicules are known for long time as they have been frequently observed at the solar limb \citep{Beckers1968}. Typical life time of the spicules is 5-15 min and upward velocity is $\sim$ 20-25 km/s. Note that the disk counterparts of spicules are called mottles \citep{Tsiropoula2012}. Recent Hinode observations with high spatial and temporal resolutions revealed spicules with different properties known as type II spicules \citep{DePontieu2007}. The type II spicules have much shorter life time of 10-150
s and higher upward velocities of 50-150 km/s than the classical spicules (now known as type I spicules). The disk counterparts of type II spicules are known as Rapid Blue/Rapid Red shifted excursions (RBEs/RREs) and were first observed by Swedish Solar Telescope \citep{Ruppe2009}. Generation mechanism for spicules is poorly understood. Classical spicules can be easily formed due to rebound shocks after photospheric pulses \citep{Hollweg1982,Murawski2010}, but type II spicules are more difficult to be excited. Recent models of emerging bipoles resulting in magnetic reconnection \citep{Moore2011,Sterling2015, Sterling2020} or in release of twist by ambipolar diffusion \citep{Martinez2017} seem to capture essential features of type II spicules, though more work is required in this regards.
Density and temperature are similar in both types of spicules, therefore the short life time
of type II spicules can be caused either due to their rapid diffusion or due to rapid heating, which may lead to their disappearance in chromospheric spectral lines. The appearance of spicules in transition region spectral lines show that type II spicules are rapidly heated \citep{Pereira2014}, though the heating mechanism is not yet completely clear. Ion-neutral collisions, Kelvin-Helmholtz instability, or both together might lead to the rapid heating \citep{Zaqarashvili2015,Kuridze2016,Martinez2017,Antolin2018}, but it is not yet fully established. Another possibility is that the spicules are quickly destroyed by some instability process. \citet{Zaqarashvili2020} suggested that the dynamic kink instability of homogeneous jet may be responsible for the disappearance of type II spicules at some height of expanding magnetic flux tubes. Here we will examine the effects of transverse inhomogeneity of flow on dynamic kink instability in both types of spicules separately.

\subsubsection{Kink instability in type I spicules}

The diameter of type I spicules varies from 300 to 1100 km \citep{Pasachoff2009}, therefore we take a mean value of 800 km, which leads to  $d=400$ km for the half width. Plasma flow may reach to 20-30 km/s so we take  $V_{0z}$=30 km/s for the velocity at the jet axis \citep{Beckers1968,Pasachoff2009}. We also assume for the external Alfv\'en speed in low corona as 200 km/s and for the density ratio of spicules and lower corona as 100. Left panels of Fig. \ref{Figure7} shows the solutions of the dispersion relation (\ref{eq:disp}) in the parameters of type I spicules. It is seen that the jets is almost fully stabilised (red lines). Only negligible instability region around $k_z d$=4.5 was found. On the other hand, hydrodynamic jets are unstable in the region of $k_z d<$ 2.4, which may correspond to significantly inclined  spicules. Right panels show the periods and the growth times of unstable harmonics vs wavelength around instability region. The period of unstable harmonics is around 23-25 s, while the growth time is 400-1000 s. The growth time of unstable harmonics is comparable to the life time of type I spicules, which is 5-15 min \citep{Beckers1968}. Hence, though the instability is weak in type I spicules, it may still destroy the structure over the observed life time.

\subsubsection{Kink instability in type II spicules}

Type II spicules are generally thinner than type I spicules with a diameter of $<200$ km \citep{DePontieu2007}, therefore we take $d=100$ km for the half width. Plasma flow may reach to 50-150 km/s so we take  $V_{0z}$=100 km/s for the velocity at the jet axis. We assume the same external Alfv\'en speed and the density ratio as for the case of the type I spicules i. e. Alfv\'en speed of  200 km/s and the density ratio of spicules and lower corona equal to 100.
Left panels of Fig. \ref{Figure8} show the solutions of the dispersion relation (\ref{eq:disp}) in these parameters. It is seen that the jets becomes unstable for the wavenumbers of $k_z d$=0.5-2.5, which corresponds to the wavelengths of 250-1400 km. Hydrodynamic jets are again unstable in the region of $k_z d<$ 2.4. Right panels show the periods and the growth times of unstable harmonics vs wavelength in instability region. The period of unstable harmonics is around 5-30 s, while the growth time is 5-60 s. The growth time of unstable harmonics is comparable to the life time of type II spicules, which is 10-150 s \citep{DePontieu2007}, Hence, the instability may destroy the type II spicules over the observed life time.

\subsubsection{Ion-neutral collision effects on the kink instability in spicules} 

Chromospheric plasma is partially ionised, therefore the collision between ions and neutral atoms may have influence on the kink instability in spicules. Since our analysis concerns only fully ionised plasma, it is of importance to estimate the effects of ion-neutral collisions. \citet{Kuridze2016} estimated the heating time due to ion-neutral collision effects as 
\begin{equation}\label{eq:neutral}
t_{heat} \sim \frac{\beta \delta_{in} D^2}{V^2_A}\frac{1-\xi_n}{\xi^2_n},
\end{equation}
where $\beta=8\pi p/B^2_z$ is the plasma beta, $D$ is the spatial scale of perturbations, $\delta_{in}$ is the ion-neutral collision frequency, $\xi_n$ is the ratio of neutral to total particle density. Let us estimate the heating time for type II spicules, taking the spatial scale of unstable harmonics as $D$=400 km (see previous subsection). We take the following values for other parameters as $\delta_{in}$= 10$^3$ Hz, $\beta=0.1$, $V_A=$ 100 km/s and $\xi_n=0,5$ \citep{Kuridze2016}. In this case, the heating time is estimated as 3.2 10$^3$ s, which is by the two orders of magnitude longer than  the growth time of kink instability. Therefore, ion-neutral collision effects are negligible at the initial stage of instability. However, when the kink instability is fully developed then the energy will be transferred to smaller scales, which will decrease the heating time. Therefore, ion-neutral collision effect may have influence on plasma dynamics only on later stage of kink instability. This is not the scope of the present paper and will be studied in the nearest future.

\subsection{Numerical simulations}

\begin{figure*}[]
   \centering
   \includegraphics[width=18cm]{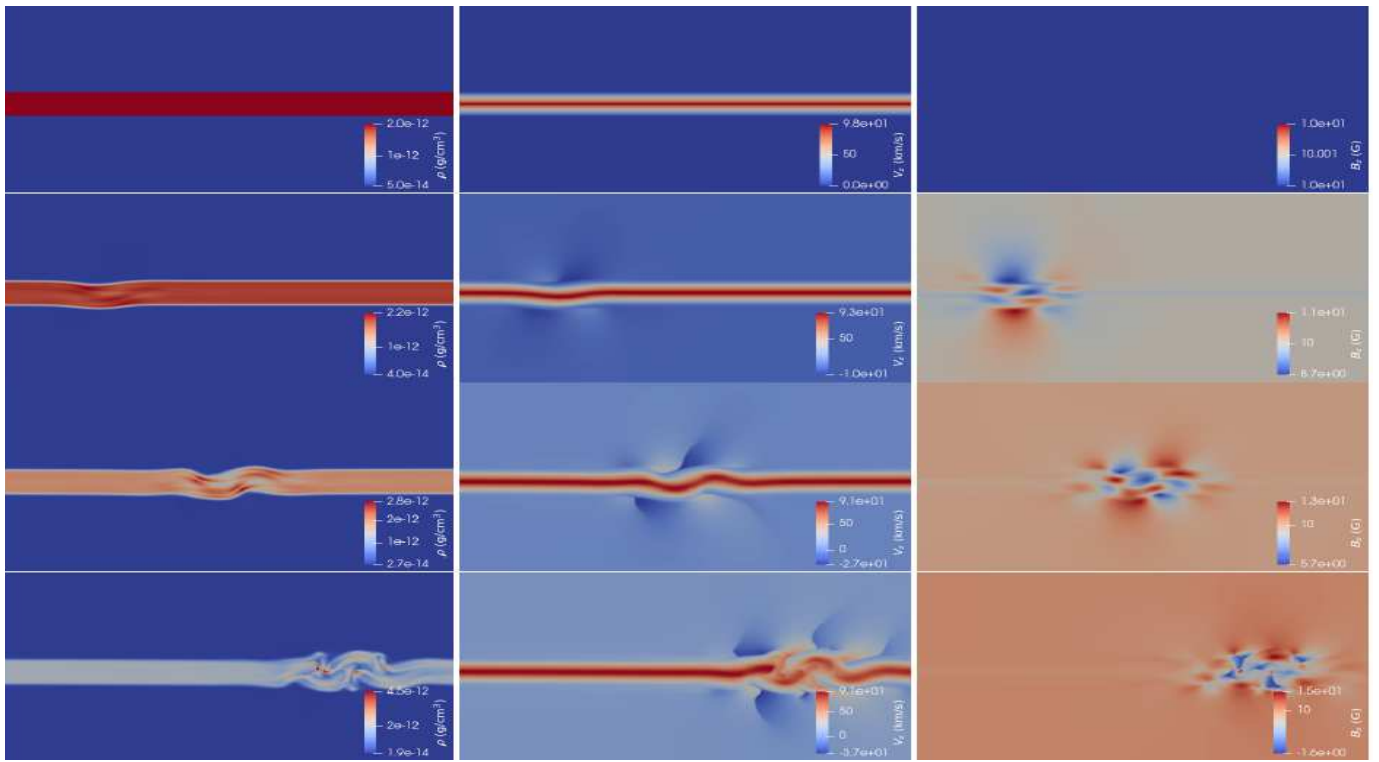}
    \caption{MHD simulation of a magnetised triangular jet in conditions of type~II spicules. The halfwidth of the jet is 100~km and the total length of the simulated jet is about 7~Mm. The initial transverse perturbation has an amplitude of $\sim 10\ \mathrm{km/s}$ (corresponding to 10\% of the flow speed at the jet centre) and a spatial scale of around 600~km. Shown are the mass density $\rho$ (left panels), the flow velocity $V_z$ (middle panels) and the $z$-component of the magnetic field $B_z$ (right panels) at consecutive snapshots, each 15~s apart (upper panels correspond to the initial set up). A pronounced kink is developing around 30~s after initial perturbation turning into an instability after $\approx 45~\mathrm{s}$. Corresponding movies can be found online.} 
   \label{fig:simulation}
\end{figure*}

The analytical model considered here is obviously simplified. The solutions are linear, therefore we only model the initial linear stage of the instability, while it is also of importance to see the full development of the instability, i.e. what is happening to the jets later on. Additionally, the absence of magnetic field inside the jet in considered analytical model is not good approximation for spicules. Therefore, numerical simulations should be invoked to test the analytical results.

In this section we present results of 2-D numerical simulations obtained with the code PLUTO, e.g. \citet{Mignone2007}, accompanying our discussion on antisymmetric kink modes in Sect.~\ref{sec:inhomogeneous-jet} in a uniform background magnetic field. The code is employed to solve the ideal MHD equations in conjuction with the thermal ideal equation of state for closure. For the flux computation an approximate Riemann solver, i.e. Harten, Lax, Van Leer (HLL) has been used. A more detailed and systematic numerical analysis is under way and will be discussed in a separate paper.

In difference with the analytical solution (with no magnetic field inside the jet), here we consider the homogeneous magnetic field inside and outside the jet with the same strength of $B_z=$ 10 G. For our setup we chose typical parameters applicable to solar type~II spicules. The jet's half width was set to $d = 100\ \mathrm{km}$ and its speed at the axis to $V_{0z} = 100\ \mathrm{km/s}$. The inner temperature is set to $10^4 \mathrm{K}$. The mass density $\rho_0$ was chosen such that a pressure equilibrium at the jet boundary is maintained, thus depending on the external mass density of $\rho_\mathrm{e} = 5 \times 10^{-14}\ \mathrm{g/cm^3}$ and external temperature of  $T_\mathrm{e} = 4 \times 10^5\ \mathrm{K}$, which means $\rho_0/\rho_\mathrm{e} = 40$. We used a cartesian grid with cell sizes of $\Delta x = \Delta z = 2\ \mathrm{km}$, corresponding to $1/50$ of the jet's half width, to easily capture the turbulent evolution of the plasma flow in the non-linear regime. The initial temporal resolution was set to $5 \times 10^{-4}\ \mathrm{s}$ and let adjust according to a maximum CFL number of 0.2. Neumann zero-gradient outflow boundary conditions have been applied on the right as well as on top and on bottom sides while an open inflow condition is specified on the left boundary.

We show that the jet becomes unstable to kink modes, when subjected to transverse antisymmetric perturbations as discussed in Sect.~\ref{sec:inhomogeneous-jet}. The perturbation amplitude of the transverse velocity was set to $\sim 10\%$ of the flow speed $V_{0z}$ and the perturbation FWHM $2 \sqrt{2 \ln 2} \sigma$ of the initial Gaussian pulse $g(z) = (\sqrt{2\pi} \sigma)^{-1} \exp\{-1/2 \cdot [(z-\mu)/\sigma]^2\}$, centered around the location $\mu$ (where the initial pulse is seeded), is related to typical sizes of granulation cells. We scanned the dispersion relation, Fig.~(3), for various wave numbers and indeed found a strong instability in the depicted regime. The simulation presented below is based on an initial perturbation with $k_z d = 1$, i.e. to the spatial scale of 628~km.

Fig. \ref{fig:simulation} shows the dynamics of the plasma density, longitudinal components of velocity and magnetic field in the jet and surroundings at different times. A significant displacement of the jet axis is already seen 15~s following the initial perturbation (top-most panels). After 45~s (bottom line) the transverse displacement already shows nonlinear character, therefore the growth time of perturbations can be estimated around 30~s, which is comparable to the analytical growth time (Fig. \ref{Figure8}). Around 45~s after the initial perturbation, the jet is almost destroyed locally. Hence, in less than 1 min after the initial excitation, the transverse velocity pulse transformed into a fully developed local instability. If one initially excites a harmonic wave instead of local pulse, then the instability will be developed along the whole wave train which will collapse the jet (numerical simulations will be presented in a separate paper). This fairly agrees with the life time of type II spicules, therefore the spicules could be destroyed by the kink instability. 

The propagation speed of the excited kink pulse is estimated as $\approx 54\ \mathrm{km/s}$ as measured in the inertial frame. The instability keeps growing until the jet is finally destroyed. The simulations show that the development of the instability critically depends on the flow speed of the jet as expected: at a lower $V_{0z}$ of $30\ \mathrm{km/s}$, the initially generated kink wave forms only a narrow, low amplitude pulse that is propagating with the flow without tearing down the underlying jet.

\section{Conclusions}
The stability of triangular jets sandwiched between two magnetic tubes/slabs in the solar atmosphere was studied. Dispersion equation governing the antisymmetric kink perturbations was obtained, which was solved analytically and numerically for different cases of Alfv\'en Mach number, $M_A$, and the ratio of internal and external densities, $\rho_0/\rho_e$. It has been shown that triangular jets are unstable to the dynamic kink instability depending on  Alfv\'en Mach number and the density ratio. For example, jets with $\rho_0/\rho_e$=1 are unstable in super  Alfv\'enic regime ($M^{-1}_A<1$), while denser jets with $\rho_0/\rho_e$=10 are unstable also in sub Alfv\'enic regime ($M^{-1}_A<2.5$). The jets flowing with an angle to the external magnetic field become unstable well below the thresholds. The results were applied to the conditions of type I and type II spicules in the solar atmosphere. The instability growth time in the conditions of type I spicules is estimated as 6-15 min depending on the perturbation wavelength, which corresponds to the life time of the spicules. The instability growth time in the conditions of type II spicules is estimated as 5-60 s depending on the perturbation wavelength, which also corresponds to the life time of these spicules. Numerical simulations of fully nonlinear MHD equations show that the initial kink pulse leads to the local breakdown of the jet in less than a minute. Consequently, the kink instability may lead to the observed short life time of type II spicules. However, simple consideration of jet and environment in this paper does not fully correspond to the real conditions of the solar atmosphere. Therefore, more work is required in future. More detailed numerical simulation of the process is currently under way.


\begin{acknowledgements}
The work was funded by the Austrian Science Fund (FWF, projects P30695-N27 and I 3955-N27) and by the Deutsche Forschungsgemeinschaft (DFG, German Research 
Foundation) - Projektnummer 407727365. SL and PG acknowledge the support of the project VEGA 2/0048/20. Acknowledgment is also given to the developers of the code PLUTO we have been using for performing the simulation presented in this paper. The authors thank the anonymous referee for stimulating comments, which led to improve the paper.
\end{acknowledgements}

%
%

\bibliographystyle{aa} 
\bibliography{ms} 

%
%
%
%
%
%
%
%
%

\end{document}